%% file: main.tex
\documentclass[11pt]{article}
\usepackage[a4paper,margin=1in]{geometry}

\usepackage[ruled,vlined,linesnumbered]{algorithm2e}
\usepackage[T1]{fontenc}
\usepackage[utf8]{inputenc}
\usepackage{microtype}
\usepackage{amsmath,amssymb,mathtools,amsthm}
\usepackage{thmtools,thm-restate}
\usepackage{bm}
\usepackage{xspace}
\usepackage{enumitem}
\usepackage[hidelinks]{hyperref}
\usepackage[capitalize,nameinlink]{cleveref}
\usepackage{booktabs}
\usepackage{nicefrac}
\usepackage{comment}
\usepackage{xcolor}

\numberwithin{equation}{section}

\newcommand{\Otil}{{\widetilde{O}}}
\newcommand{\poly}{\mathrm{poly}}

\newcommand{\opt}{\mathrm{OPT}}
\newcommand{\cut}{\delta}
\newcommand{\val}{\mathrm{val}}

\newcommand{\mindeg}[1]{{\mathrm{deg}_{\min}}(#1)}

\newtheorem{theorem}{Theorem}[section]
\newtheorem{lemma}[theorem]{Lemma}

\newtheorem{proposition}[theorem]{Proposition}
\newtheorem{corollary}[theorem]{Corollary}

\newtheorem{definition}[theorem]{Definition}
\newtheorem{fact}[theorem]{Fact}
\newtheorem{remark}[theorem]{Remark}

\title{Sub-$n^k$ Deterministic algorithm for minimum $k$-way cut in simple graphs}

\author{Mohit Daga\thanks{KTH - Royal Institute of Technology, Stockholm, Sweden. \texttt{mdaga@kth.se}}}
\date{\today}

\begin{document}
	
	\begin{titlepage}
		\maketitle
		\pagenumbering{roman}
		
		\begin{abstract}
			We present a \emph{deterministic exact algorithm} for the \emph{minimum $k$-cut problem} on simple graphs.
			Our approach combines the \emph{principal sequence of partitions (PSP)}, derived canonically from ideal loads, with a single level of \emph{Kawarabayashi--Thorup (KT)} contractions at the critical PSP threshold~$\lambda_j$.
			Let $j$ be the smallest index with $\kappa(P_j)\ge k$ and $R := k - \kappa(P_{j-1})$.
			We prove a structural decomposition theorem showing that an optimal $k$-cut can be expressed as the level-$(j\!-\!1)$ boundary $A_{\le j-1}$ together with exactly $(R-r)$ \emph{non-trivial} internal cuts of value at most~$\lambda_j$ and $r$ \emph{singleton isolations} (``islands'') inside the parts of~$P_{j-1}$.
			At this level, KT contractions yield kernels of total size $\widetilde{O}(n / \lambda_j)$, and from them we build a \emph{canonical border family}~$\mathcal{B}$ of the same order that deterministically covers all optimal refinement choices.
			Branching only over~$\mathcal{B}$ (and also including an explicit ``island'' branch) gives total running time
			\[
			T(n,m,k) = \widetilde{O}\left(\mathrm{poly}(m)+\Bigl(\tfrac{n}{\lambda_j}+n^{\omega/3}\Bigr)^{R}\right),
			\]
			where $\omega < 2.373$ is the matrix multiplication exponent.
			In particular, if $\lambda_j \ge n^{\varepsilon}$ for some constant $\varepsilon > 0$, we obtain a \emph{deterministic sub-$n^k$-time algorithm}, running in $n^{(1-\varepsilon)(k-1)+o(k)}$ time.
			Finally, combining our PSP$\times$KT framework with a small–$\lambda$ exact subroutine via a simple meta-reduction yields a deterministic $n^{c k+O(1)}$ algorithm for $c = \max\{ t/(t+1), \omega/3 \} < 1$, aligning with the exponent in the randomized bound of He--Li (STOC~2022) under the assumed subroutine.
		\end{abstract}
		
		\newpage
		\setcounter{tocdepth}{2}
		\tableofcontents
	\end{titlepage}
	
	\newpage
	\pagenumbering{arabic}
	
	\section{Introduction}
	
	The \emph{minimum $k$-cut problem} is a fundamental graph partitioning task: given an undirected graph $G=(V,E)$ and an integer $k\ge2$, the goal is to delete the fewest edges so that the graph breaks into at least $k$ connected components.
	Let $\lambda_k$ denote the minimum size of such an edge set.
	When $k=2$, the problem reduces to the well-known global minimum cut, solvable in near-linear time; for larger $k$, however, the combinatorial and algorithmic structure becomes substantially richer.
	
	\paragraph{Classical background.}
	For constant $k$, the first polynomial-time algorithm was obtained by Goldschmidt and Hochbaum~\cite{GH94}, whose approach runs in $n^{O(k^2)}$ time.
	A major simplification came with the contraction framework of Karger and Stein~\cite{karger-stien}, which gave a randomized $\widetilde{O}(n^{2k-2})$ algorithm.
	Thorup~\cite{Thorup08} later matched this deterministically using tree packings, achieving $\widetilde{O}(n^{2k})$ time.
	Over the following two decades, several refinements~\cite{GLL18,GLL19,Li19} culminated in the $\widetilde{O}(n^{k})$ algorithm of Gupta, Harris, Lee, and Li~\cite{GHLL20}, whose analysis showed that Karger–Stein’s recursion is essentially optimal under the reduction from the \emph{max-weight $k$-clique} problem.
	Under this reduction, no $\widetilde{O}(n^{k-\delta})$ algorithm is possible unless max-weight $k$-clique admits an equally fast solution.
	
	\paragraph{The unweighted gap.}
	For unweighted graphs, however, the picture is more nuanced.
	The reduction in~\cite{GHLL20} does not preclude faster algorithms, and the best known upper bound for unweighted $k$-clique---and therefore a plausible target for unweighted $k$-cut---is $n^{(\omega/3)k+O(1)}$, where $\omega<2.373$ is the matrix multiplication exponent~\cite{AW21}.
	This raised a natural question: can the unweighted minimum $k$-cut be solved in sub-$n^k$ time?
	Recently, He and Li~\cite{HeLi22} resolved this question \emph{for randomized algorithms}, presenting an exact algorithm for simple graphs running in $n^{(\omega/3)k+O(1)}$ time by integrating random sampling with local sparsification.
	Their result established the first truly sub-$n^k$ upper bound but relied essentially on randomization.
	Whether a comparable bound could be achieved \emph{deterministically} remained open.
	
	\paragraph{Our Contribution.}
	We give the first \emph{deterministic exact algorithm} for minimum $k$-cut on simple graphs that runs in \emph{sub-$n^k$ time}.
	Our approach unifies two canonical frameworks: the \emph{principal sequence of partitions (PSP)}---derived deterministically from the ideal-load dual of the cut LP---and a single level of \emph{Kawarabayashi--Thorup (KT)} contractions at the critical PSP threshold~$\lambda_j$.
	Let $j$ be the smallest index with $\kappa(P_j) \ge k$ and set $R := k - \kappa(P_{j-1})$.
	We prove that every optimal $k$-cut can be represented as
	\[
	F^\star = A_{\le j-1} \,\cup\, \bigcup_{q=1}^{R-r} B_q \,\cup\, \bigcup_{p=1}^{r} I_p,
	\]
	where each $B_q$ is a non-trivial internal border of value $\le \lambda_j$ inside a part of $P_{j-1}$ and each $I_p$ is a singleton (``island'').
	At level $\lambda_j$, KT contractions yield kernels of total size $\Otil(n/\lambda_j)$, from which we construct a \emph{canonical border family} $\mathcal{B}$ of size $\Otil(n/\lambda_j)$ that deterministically covers all possible optimal refinements.
	Branching only over $\mathcal{B}$ (plus an explicit ``stop-and-island'' option) yields
	\[
	T(n,m,k) = \Otil\!\left(\poly(m)+\Bigl(\tfrac{n}{\lambda_j}+n^{\omega/3}\Bigr)^R\right).
	\]
	In particular, if $\lambda_j \ge n^{\varepsilon}$ for some constant $\varepsilon>0$, we obtain a deterministic sub-$n^k$ algorithm running in $n^{(1-\varepsilon)(k-1)+o(k)}$ time. For smaller values, we use the result of Lokshtanov, Saurabh, and Surianarayanan \cite{LSS20}, who showed an algorithm for exact minimum k-cut that runs in time $\lambda^{O(k)}n^{O(1)}$.
	
	\paragraph{Deterministic $n^{ck}$ Meta-Theorem.}
	Assuming access to a deterministic exact routine for small~$\lambda$, a simple reduction yields a deterministic
	\[
	T(n,k) = O\!\bigl(n^{c k + O(1)}\bigr)
	\]
	algorithm for $c = \max\{t/(t+1),\, \omega/3\} < 1$, aligning with the exponent of the randomized algorithm of He--Li~\cite{HeLi22} under the same $k$-clique barrier.
	
	\paragraph{Perspective.}
	Our results highlight the algorithmic power of the PSP/ideal-load structure as a canonical deterministic backbone for global cut problems and demonstrate that a single KT contraction level suffices to reach the sub-$n^k$ regime.
	
	% ============================================================
	% 2. Preliminaries: Notation
	% ============================================================
	\section{Preliminaries}
	\label{sec:notation}
	Let $G=(V,E)$ be a simple undirected graph; $n:=|V|$, $m:=|E|$. For $S\subseteq V$, $\cut(S)$ denotes the set of edges with exactly one endpoint in $S$. For  \(A,B\subseteq V\) and a graph \(H\), define $\delta_H(A,B):=\{\{u,v\}\in E(H): (u\in A\setminus B\ \wedge\ v\in B\setminus A)$. For a partition $P$ of $V$, let $\kappa(P)$ denote its number of parts. We use unit capacities unless stated otherwise; the value of a cut $F\subseteq E$ is $\val(F)=|F|$. A \emph{$k$-cut} is a set $F\subseteq E$ whose removal yields at least $k$ connected components. We write $\Otil(\cdot)$ for $O(\cdot\,\poly\log n)$ and use $\omega<2.373$ for the matrix multiplication exponent.
	
	% ============================================================
	% 3. PSP / Ideal-Loads Backbone
	% ============================================================
	\input{03_psp-ideal-load-revised}

	% 4. One-Level KT Kernels at \lambda_j
	% ============================================================
	\input{04_kt_kernels}
	% ==========================================================================================================
	% 5. Algorithmic Framework
	% ============================================================
	\input{05_algorithm}
	% ============================================================
	% 6. Correctness and Running Time
	% ============================================================
	\input{06_correctness}
	
	% ============================================================
	% 7. Deterministic n^{ck} via PSP×KT + small–\lambda routine
	% ============================================================
	\input{07_last_meta}

	\bibliographystyle{alpha}
	\bibliography{references}
	\appendix
	\section{Omitted Proofs}
	\input{omitted_proofs}
\end{document}

%% file: 03_psp-ideal-load-revised.tex
\subsection{PSP / Ideal-Loads Backbone}
\label{sec:psp}\paragraph{Setting and notation.}
All graphs are simple and uncapacitated unless explicitly stated. For a partition $P$ of $V$,
write $\kappa(P)$ for the number of parts, and $G[P]$ for the graph restricted to each part.
All cut values and degrees inside a part $C\in P$ are with respect to the induced subgraph
$G[C]$ unless noted otherwise.

\begin{definition}[PSP and strengths]\label{def:psp}
A \emph{principal sequence of partitions (PSP)} is a chain
$P_0 \succ P_1 \succ \cdots \succ P_t$ with strictly increasing \emph{strengths}
$0<\lambda_1<\cdots<\lambda_t$
such that $P_i$ is obtained from $P_{i-1}$ by deleting a canonical edge set $A_i$ whose total value equals
\[
\val(A_i)\;=\;\bigl(\kappa(P_i)-\kappa(P_{i-1})\bigr)\cdot \lambda_i.
\]
We write $A_{\le i}:=\bigcup_{q\le i}A_q$ and $B_i:=A_i\setminus A_{\le i-1}$.
\end{definition}

\subsection{Ideal loads: definition and key properties}
The load formalism we use mirrors Thorup's ideal relative loads for tree packings and their
duality to partitions \cite{Thorup2007}; we summarize the facts we need. Let
$\Phi(G):=\min_{P}\frac{|E(G/P)|}{|P|-1}$ be the \emph{partition value} of $G$; by
Tutte/Nash-Williams, $\Phi(G)$ equals the maximum packing value of spanning trees. When defining $\Phi$ and constructing ideal loads, we may assume that each optimal partition has connected parts. Th
Define the \emph{ideal relative loads} $\ell^\star:E\to\mathbb{R}_{>0}$ recursively as follows
(\cite{Thorup2007}):
\begin{enumerate}
\item Let $P^\star$ be a partition of $V$ attaining $\Phi(G)$.
Set $\ell^\star(e)=1/\Phi(G)$ for all $e\in E(G/P^\star)$ (edges crossing distinct parts).
\item For each part $S\in P^\star$, recurse on $G[S]$ with its own $\Phi(G[S])$, defining
$\ell^\star(\cdot)$ for edges internal to $S$.
\end{enumerate}

\paragraph{Greedy tree packings and concrete loads.}
Let $T_1,\dots,T_t$ be a \emph{greedy} tree packing (each $T_i$ is an MST w.r.t.\ the loads
induced by $\{T_1,\dots,T_{i-1}\}$). Let $L_{\mathcal T}(e)$ be the number of packed trees
containing $e$ and $\ell_{\mathcal T}(e):=L_{\mathcal T}(e)/t$ the \emph{concrete} relative load.

\begin{fact}[Ideal-loads toolkit]\label{fact:ideal-loads-toolkit}
Let $\Phi(G):=\min_{P}\frac{|E(G/P)|}{|P|-1}$ be the minimum partition value.
There exists a vector of \emph{ideal relative loads} $\ell^\star:E(G)\to\mathbb{R}_{>0}$ and
a distribution $\Pi$ over spanning trees such that $\Pr_{T\sim\Pi}[e\in T]=\ell^\star(e)$ for all $e$, and the construction of $\ell^\star$ proceeds \emph{recursively} by: taking an optimal partition $P^\star$ with $\Phi(G)$ and then recursing on each induced subgraph $G[S]$ for $S\in P^\star$. Consequently:
\begin{enumerate}
\item[(i)] (\emph{Tutte--Nash-Williams duality}) $\max_{\mathcal{T}}\text{pack\_val}(\mathcal{T})=\min_{ P}\text{part\_val}(P)=\Phi(G)$, and
\(\tfrac{1}{2}\lambda(G)<\Phi(G)\le \lambda(G)\). Here $\mathcal{T}$ is a family of spanning-trees and $P$ is a partition of vertex sets.% Thorup Lemma 3
\item[(ii)] (\emph{Plateaus}) The function $\ell^\star$ is \emph{piecewise constant}: there are strictly decreasing values $L_1>L_2>\cdots$ so that on each recursion level $i$ the set $B_i:=\{e:\ell^\star(e)=L_i\}$ is a disjoint layer; deleting $A_{\le i}:=\bigcup_{q\le i}B_q$ refines the current partition. Plateaus strictly decrease when recursing into subgraphs.
\item[(iii)] (\emph{Laminarity}) For any fixed level, the family of boundaries exposed by deleting that level’s edges refines parts by splits only; by uncrossing these boundaries form a \emph{laminar} family. Moreover, peeling pendant blocks yields an \emph{independent} subfamily whose boundaries are edge-disjoint in the original graph.
\end{enumerate}
All statements above follow from the ideal-load recursion and the cut/partition duality; see \cite[§2–§4]{Thorup2007}. % cites Thorup
\end{fact}
\begin{proposition}[Greedy loads uniformly approximate ideal loads {\normalfont\cite[Proposition 16]{Thorup2007}}]
\label{prop:uniform-approx}
Let $\varepsilon\in(0,2)$ and suppose $\mathcal T$ is a greedy packing of
\[
t\;\ge\;\frac{6\,\lambda(G)\,\ln m}{\varepsilon^2}
\]
trees. Then for every $e\in E(G)$,
\[
\bigl|\ell_{\mathcal T}(e)-\ell^\star(e)\bigr|\;\le\;\frac{\varepsilon}{\lambda(G)}.
\]
\end{proposition}
\noindent
\section{PSP from ideal loads}
\label{sec:psp-from-loads}

\paragraph{Conceptual reconstruction via load plateaus.}
Let $\ell^\star:E\to\mathbb{R}_{>0}$ be the ideal relative loads from
\cref{fact:ideal-loads-toolkit}.
Let the distinct values taken by $\ell^\star$ be
\[
L_1 > L_2 > \cdots > L_t > 0.
\]
For each $i$, define the \emph{plateau} and its prefix
\[
B_i := \{e\in E : \ell^\star(e)=L_i\},
\qquad
A_{\le i} := \bigcup_{q\le i} B_q.
\]
Let $P_i$ be the partition of $V$ into connected components of $G-A_{\le i}$ (and $P_0:=\{V\}$).

\begin{theorem}[PSP from ideal loads (with partition strengths)]
\label{thm:psp-from-loads}
Let $P_0 \succ P_1 \succ \cdots \succ P_t$ be the chain induced by the load plateaus as above.
For each $i$ define the \emph{partition strength}
\[
\lambda_i \;:=\; \frac{1}{L_i}.
\]
Then $(P_0\succ \cdots \succ P_t)$ is a PSP in the sense of \cref{def:psp} with levels $A_i:=B_i$
and strictly increasing strengths $0<\lambda_1<\cdots<\lambda_t$; in particular, for every $i$,
\[
\val(A_i)
\;=\;
\bigl(\kappa(P_i)-\kappa(P_{i-1})\bigr)\cdot \lambda_i.
\]
Moreover, each $\lambda_i$ equals the partition value $\Phi(H)$ of the induced subgraphs $H$
that are first split at level $i$ in the ideal-load recursion.
\end{theorem}

\begin{proof}
Fix a level $i$. By the recursive definition of ideal loads (see \cref{fact:ideal-loads-toolkit}),
each edge $e$ with $\ell^\star(e)=L_i$ is a crossing edge of some optimal partition $P^\star$ of
an induced subgraph $H$ encountered by the recursion, and for that $H$ we have
\[
L_i \;=\; \frac{1}{\Phi(H)}
\qquad\text{and hence}\qquad
\lambda_i \;=\; \Phi(H).
\]
For such an $H$ and optimal $P^\star$ attaining $\Phi(H)$,
\[
|E(H/P^\star)|
\;=\;
(|P^\star|-1)\,\Phi(H)
\;=\;
(|P^\star|-1)\,\lambda_i.
\]
Deleting $E(H/P^\star)$ increases the number of connected components inside $H$ by exactly
$|P^\star|-1$.

Now delete \emph{all} edges in $B_i$ globally. At this recursion level the induced subgraphs $H$
being split have disjoint vertex sets (they are parts created by earlier recursion), so their
contributions add:
the total increase in the number of global parts is
$\kappa(P_i)-\kappa(P_{i-1})$, and the total number of edges deleted is $\val(B_i)$.
Since every contributing $H$ at this level has the same $\Phi(H)=\lambda_i$ (equivalently, the
same load $L_i$), summing the displayed identity over all such $H$ yields
\[
\val(B_i)
\;=\;
\bigl(\kappa(P_i)-\kappa(P_{i-1})\bigr)\lambda_i.
\]
Taking $A_i:=B_i$ gives the PSP identity.

Finally, plateau values strictly decrease by \cref{fact:ideal-loads-toolkit}(ii), so
$\lambda_i=1/L_i$ strictly increases.
\end{proof}

\begin{remark}[Min-cuts vs.\ partition strengths]
For any induced subgraph $H$, every cut is a $2$-part partition, hence $\Phi(H)\le \lambda(H)$.
Moreover \cref{fact:ideal-loads-toolkit}(i) gives $\tfrac12\lambda(H)<\Phi(H)\le \lambda(H)$.
Thus, whenever a part $C$ is first split at level $i$, its minimum cut value satisfies
\[
\lambda_i \;\le\; \lambda(G[C]) \;<\; 2\lambda_i.
\]
In particular, if one tries to interpret the PSP strength at level $i$ as a minimum cut value,
the exact PSP identity above need not hold; in general one can only guarantee an inequality.
\end{remark}

\paragraph{Deterministic computation (loads $\to$ PSP).}
We now give concrete steps to \emph{compute} the PSP deterministically from a sufficiently accurate
greedy tree packing.

\begin{algorithm}[H]
\caption{\textsc{LoadSweep-PSP}}
\label{alg:loadsweep}
\KwIn{Graph $G=(V,E)$}
\KwOut{PSP $(P_0 \succ \cdots \succ P_t)$ and edge sets $(A_{\le i})_{i \ge 0}$}

Build a greedy tree packing with
\[
t\;\ge\;\frac{6\,\lambda(G)\,\ln m}{\varepsilon^2}
\]
trees (for $\varepsilon$ fixed below), and let $\ell_{\mathcal{T}}(e)$ be the resulting concrete loads. \\

Sort $E$ by $\ell_{\mathcal{T}}(e)$ in decreasing order and scan to create \emph{robust plateaus}:
start a new level whenever the gap between consecutive distinct load values exceeds $2\varepsilon/\lambda(G)$.
Let these buckets be $\widehat{B}_1, \widehat{B}_2, \ldots$ and define
$\widehat{A}_{\le i} = \bigcup_{q \le i} \widehat{B}_q$. \\

Return the partitions $\widehat{P}_i$ induced by deleting $\widehat{A}_{\le i}$.
\end{algorithm}

\begin{lemma}[Correctness of \textsc{LoadSweep-PSP}]
\label{lem:loadsweep-correct}
Fix $\varepsilon\le \lambda(G)/(4m^2)$. Algorithm~\ref{alg:loadsweep} returns the canonical PSP induced by
the ideal-load plateaus.
\end{lemma}
\begin{proof}
By \cref{prop:uniform-approx}, for the chosen sample size
\(t \ge 6\lambda(G)\ln m/\varepsilon^2\), we have the uniform bound
\begin{equation}
\label{eq:unif-approx}
\bigl|\ell_{\mathcal T}(e)-\ell^\star(e)\bigr|
\;\le\; \frac{\varepsilon}{\lambda(G)}
\qquad\text{for all } e\in E.
\end{equation}

\paragraph{Minimum separation between distinct ideal plateaus.}
Each ideal plateau value is of the form
\[
L \;=\; \frac{1}{\Phi(H)} \;=\; 
\frac{|P|-1}{|E(H/P)|},
\]
for some induced subgraph \(H\) and partition \(P\).
Thus \(L\) is a rational number whose reduced denominator is at most
\(|E(H/P)| \le m\).  
Hence any two distinct plateau values \(L \neq L'\) satisfy
\begin{equation}
\label{eq:plateau-gap}
|L - L'| \;\ge\; \frac{1}{m^2}.
\end{equation}

\paragraph{Noise is below half the minimum gap.}
Assume \(\varepsilon \le \lambda(G)/(4m^2)\).  
Then the scanning threshold obeys
\begin{equation}
\label{eq:scan-threshold}
\frac{2\varepsilon}{\lambda(G)} \;\le\; \frac{1}{2m^2}.
\end{equation}
Combining \eqref{eq:plateau-gap} and \eqref{eq:scan-threshold},
\[
\frac{2\varepsilon}{\lambda(G)} 
\;<\; \frac{|L-L'|}{2}
\qquad\text{for all distinct plateau values } L\neq L'.
\]

\paragraph{The scan cannot merge distinct plateaus.}
Let edges \(e \in B_i\) and \(e' \in B_j\) belong to two distinct ideal
plateaus with values \(L_i > L_j\).  
Using \eqref{eq:unif-approx},
\[
\ell_{\mathcal T}(e)
\;\ge\; L_i - \frac{\varepsilon}{\lambda(G)},
\qquad
\ell_{\mathcal T}(e')
\;\le\; L_j + \frac{\varepsilon}{\lambda(G)}.
\]
Therefore
\begin{equation}
\label{eq:empirical-diff}
\ell_{\mathcal T}(e) - \ell_{\mathcal T}(e')
\;\ge\;
(L_i - L_j) - \frac{2\varepsilon}{\lambda(G)}.
\end{equation}
Using \eqref{eq:plateau-gap} and \eqref{eq:scan-threshold},
\[
\ell_{\mathcal T}(e) - \ell_{\mathcal T}(e')
\;\ge\;
\frac{1}{m^2} - \frac{1}{2m^2}
\;=\;
\frac{1}{2m^2}
\;>\;
\frac{2\varepsilon}{\lambda(G)}.
\]
Thus the load-scan must place a bucket boundary between the two plateaus,
so no merging is possible.

\paragraph{The scan cannot split a single plateau.}
If \(e,e' \in B_i\), then both satisfy
\[
\ell_{\mathcal T}(e),\, \ell_{\mathcal T}(e')
\in 
\left[
L_i - \frac{\varepsilon}{\lambda(G)},\;
L_i + \frac{\varepsilon}{\lambda(G)}
\right].
\]
Thus
\[
|\ell_{\mathcal T}(e) - \ell_{\mathcal T}(e')|
\;\le\; \frac{2\varepsilon}{\lambda(G)},
\]
so the scan cannot introduce a boundary within a true plateau.

\paragraph{Conclusion.}
From the above two facts, each estimated bucket \(\widehat{B}_i\)
contains exactly the edges of the true plateau \(B_i\).  
Hence
\[
\widehat{B}_i = B_i
\qquad\text{for all } i.
\]
Since the partitions \(P_i\) and \(\widehat{P}_i\) are defined by contracting
the prefix of buckets, it follows that
\[
\widehat{P}_i = P_i
\qquad\text{for all } i.
\]
Therefore LoadSweep--PSP reconstructs the exact
\end{proof}

\begin{comment}
    
\begin{proof}
By \cref{prop:uniform-approx}, for the chosen $t$ we have
$|\ell_{\mathcal T}(e)-\ell^\star(e)|\le \varepsilon/\lambda(G)$ for every $e\in E$.

Each ideal plateau value is of the form $L=1/\Phi(H)=(|P|-1)/|E(H/P)|$ for some induced subgraph $H$
and partition $P$, hence a rational number whose reduced denominator is at most $|E(H/P)|\le m$.
Therefore any two distinct plateau values differ by at least $1/m^2$.
With $\varepsilon\le \lambda(G)/(4m^2)$,e the scanning threshold
$2\varepsilon/\lambda(G)\le 1/(2m^2)$ is smaller than half the minimum gap between distinct ideal plateaus.
Thus the scan cannot merge two different ideal plateaus and cannot split a single ideal plateau.
Hence each bucket $\widehat{B}_i$ equals the true plateau $B_i$, and the induced partitions match:
$\widehat{P}_i=P_i$ for all $i$.
\end{proof}
\end{comment}

%=========
%=========
%=========

\input{03_1_psp-ideal-load-revised}

%% file: 03_1_psp-ideal-load-revised.tex
\subsection{PSP-compatibility of OPT}

\label{subsec:psp-compat}

Fix an integer $k\ge 2$ and let $P_0 \succ P_1 \succ \cdots \succ P_t$ be the canonical PSP induced by
the ideal-load plateaus from \cref{thm:psp-from-loads}. Recall that $A_{\le i}$ denotes the prefix of
deleted edges up to level $i$ and that $P_i$ is the partition into connected components of $G-A_{\le i}$.
Let
\[
j \;:=\; \min\{i : \kappa(P_i)\ge k\},
\qquad
R \;:=\; k-\kappa(P_{j-1}).
\]
Thus, starting from $P_{j-1}$ (which has $\kappa(P_{j-1})=k-R$ parts), any optimal $k$-cut must create
exactly $R$ further components \emph{inside} the parts of $P_{j-1}$.

\paragraph{Borders and islands.}
A cut inside a current part $C$ is \emph{non-trivial} if both sides contain at least two vertices.
We also allow \emph{islands}, i.e., singleton isolations: isolating $v\in C$ costs $\deg_{G[C]}(v)$.

At the $P_{j-1}$ level we will use the following canonical (integer) cut threshold:
\[
\theta_j
\;:=\;
\min\bigl\{\,|\delta_{G[C]}(S)| : C\in P_{j-1},~ S\subset C,~ 2\le |S|\le |C|-2\,\bigr\},
\]
with the convention $\theta_j=+\infty$ if the set is empty (i.e., no non-trivial internal cut exists in any
part of $P_{j-1}$).
A \emph{(level-$j$) border} is any non-trivial internal cut of value $\theta_j$ taken inside a current part
(a descendant of a $P_{j-1}$ part). By definition, there is no non-trivial internal cut cheaper than
$\theta_j$ in any $C\in P_{j-1}$, hence every border has value exactly $\theta_j$.

\begin{remark}[Relating $\theta_j$ to the PSP strength $\lambda_j$]
\label{rem:theta-vs-lambda}
Since every cut is a $2$-part partition, for any induced subgraph $H$ we have $\Phi(H)\le \lambda(H)$.
In particular, for every $C\in P_{j-1}$,
\[
\min_{\emptyset\subsetneq S\subsetneq C} |\delta_{G[C]}(S)|
\;\ge\;
\Phi(G[C])
\;\ge\;
\lambda_j,
\]
where the last inequality follows from \cref{fact:ideal-loads-toolkit}(ii) (partition value is non-decreasing
under recursion) together with \cref{thm:psp-from-loads}. Hence $\theta_j\ge \lambda_j$ whenever
$\theta_j<+\infty$. Moreover, whenever some part is first split non-trivially at level $j$, the min-cut
vs.\ partition-value sandwich in \cref{fact:ideal-loads-toolkit}(i) yields $\theta_j < 2\lambda_j$.
\end{remark}

%-----------------------------------------
%  Helper lemmas: submodularity / uncrossing
%-----------------------------------------

\begin{lemma}[Partition uncrossing on the lattice]
\label{lem:partition-uncrossing}
For any two partitions $P,Q$ of $V$,
\[
|E(G/P)| + |E(G/Q)|
\;\ge\;
|E(G/(P\wedge Q))| + |E(G/(P\vee Q))|,
\]
where $\wedge$ and $\vee$ denote the meet and join in the partition lattice (with refinement order).
\end{lemma}

\begin{proof}
Write $\mathbf{1}_P(e)$ for the indicator that the endpoints of $e$ lie in distinct parts of $P$. Then
$|E(G/P)|=\sum_{e\in E}\mathbf{1}_P(e)$. For a single edge $\{u,v\}$, a 4-case check gives
$\mathbf{1}_P(u,v)+\mathbf{1}_Q(u,v)\ge \mathbf{1}_{P\wedge Q}(u,v)+\mathbf{1}_{P\vee Q}(u,v)$.
Summing over all $e\in E$ yields the claim.
\end{proof}

\begin{lemma}[Cut uncrossing $\Rightarrow$ laminar minimum borders]
\label{lem:laminar-pendant-additivity}
Fix a graph $H$ and an integer $\theta\ge 1$. Assume $\theta$ is the minimum value of $|\delta_H(S)|$ over all non-trivial $S \subset V(H)$ whose both sides contain at least two vertices. Let $\mathcal{S}$ be any family of non-trivial vertex sets
$S\subset V(H)$ such that $|\delta_H(S)|=\theta$ for all $S\in\mathcal{S}$.
Then there exists a laminar family $\mathcal{L}$ of non-trivial sets with
$|\delta_H(S)|=\theta$ for all $S\in\mathcal{L}$ such that the cuts $\{\delta_H(S):S\in\mathcal{L}\}$
realize at least the same refinements as $\mathcal{S}$.
\end{lemma}

\begin{proof}
We use the standard cut submodularity:
for any $X,Y\subseteq V(H)$,
\[
|\delta_H(X)|+|\delta_H(Y)|
\;\ge\;
|\delta_H(X\cap Y)|+|\delta_H(X\cup Y)|.
\]
If $X$ and $Y$ cross (none of $X\cap Y,X\setminus Y,Y\setminus X$ is empty), then the right-hand side
forces at least one of $|\delta_H(X\cap Y)|,|\delta_H(X\cup Y)|$ to be $\le \theta$; by minimality of
$\theta$ among the non-trivial cuts in $H$, uncrossing can be performed without increasing cut value.
Iterating eliminates crossings and yields a laminar family.
\end{proof}

%-----------------------------------------
%  Step 1: normalize OPT to refine P_{j-1}
%-----------------------------------------

\input{03_07}

%-----------------------------------------
%  Main characterization
%-----------------------------------------

\begin{lemma}[PSP-normal form of OPT]
\label{lem:psp-normal-form}
Let $j:=\min\{i:\kappa(P_i)\ge k\}$ and $R:=k-\kappa(P_{j-1})$ for the canonical PSP from
\cref{thm:psp-from-loads}. Then there exists an optimal $k$-cut $F^\star$ of the form
\[
F^\star
\;=\;
A_{\le j-1}
~\cup~
\Bigl(\bigcup_{q=1}^{R-r} B_q\Bigr)
~\cup~
\Bigl(\bigcup_{p=1}^{r} I_p\Bigr),
\]
where $r\in\{0,1,\dots,R\}$, each $B_q$ is a non-trivial border cut taken inside some current part
(a descendant of a $P_{j-1}$-part) and satisfies $\val(B_q)=\theta_j$ (hence $\val(B_q)\le 2\lambda_j$ by
\cref{rem:theta-vs-lambda}), and each $I_p$ isolates a single vertex (so $\val(I_p)=\deg_{G[C]}(v_p)$ for
some $v_p$ in its current part $C$).
\end{lemma}

\begin{proof}
By \cref{lem:psp-prefix-normalization}, there is an optimal $k$-cut $F$ with $A_{\le j-1}\subseteq F$.
All remaining edges of $F$ lie inside parts of $P_{j-1}$. Let $r$ be the number of components of $G-F$
that are singletons \emph{created inside} the $P_{j-1}$-parts; isolate these $r$ vertices last and call the
corresponding singleton cuts $I_1,\dots,I_r$. The remaining $R-r$ component increases are achieved by
non-trivial internal splits; write the corresponding non-trivial internal cuts as $B'_1,\dots,B'_{R-r}$.

By definition of $\theta_j$, every non-trivial internal cut inside any original part $C\in P_{j-1}$ has
value at least $\theta_j$. Hence $\val(B'_q)\ge \theta_j$ for all $q$. Replacing any $B'_q$ by a border of
value exactly $\theta_j$ cannot increase the cut size and still increases $\kappa$ by $1$, so w.l.o.g.\ we may
assume each non-trivial split uses a border $B_q$ with $\val(B_q)=\theta_j$.

Putting these pieces together gives the claimed normal form
\(
F^\star = A_{\le j-1}\cup \bigcup_{q=1}^{R-r}B_q\cup\bigcup_{p=1}^{r}I_p.
\)
\end{proof}

%% file: 03_07.tex
\begin{lemma}[PSP-prefix normalization]
\label{lem:psp-prefix-normalization}
Let $j:=\min\{i:\kappa(P_i)\ge k\}$ and $R:=k-\kappa(P_{j-1})$. Then there exists an optimal $k$-cut
$F^\star$ whose component partition refines $P_{j-1}$. Equivalently,
\[
A_{\le j-1} \subseteq F^\star.
\]
\end{lemma}

\begin{proof}
Let $F$ be a minimum $k$-cut and let $Q$ be the partition of $V$ into connected components of $G-F$. Assume for contradiction that $A_{\le j-1}\nsubseteq F$, and set $S:=A_{\le j-1}\setminus F\neq\emptyset$.
Let $\widetilde P$ be the partition obtained from $P_{j-1}$ by \emph{gluing together} those parts that are
connected by edges of $S$ (equivalently: take the connected components of the graph obtained by contracting
each $C\in P_{j-1}$ to a supernode and keeping exactly the superedges induced by $S$). Then $\widetilde P$
is a coarsening of $P_{j-1}$ whose parts are connected in $G$ (each is a union of connected $P_{j-1}$-parts
linked by actual edges). Write
\[
d \;:=\; \kappa(P_{j-1})-\kappa(\widetilde P)\;\ge\;1.
\]

\smallskip\noindent
\textbf{(1) Bounding $|S|$ via loads.}
Every edge in $A_{\le j-1}$ lies on some plateau $1,\dots,j-1$, hence has ideal load at least
$L_{j-1}=1/\lambda_{j-1}$.
Also, for any partition into connected parts, a spanning tree crosses it in exactly $\kappa(\cdot)-1$ edges.
Taking expectation over the tree distribution underlying $\ell^\star$ gives
\[
\sum_{e\in E(G/P_{j-1})}\ell^\star(e)=\kappa(P_{j-1})-1,
\qquad
\sum_{e\in E(G/\widetilde P)}\ell^\star(e)=\kappa(\widetilde P)-1.
\]
Since $\widetilde P$ is a coarsening of $P_{j-1}$, we have $E(G/\widetilde P)\subseteq E(G/P_{j-1})=A_{\le j-1}$,
so the total load mass of $A_{\le j-1}$-edges that become internal under $\widetilde P$ is
\[
\sum_{e\in A_{\le j-1}\setminus E(G/\widetilde P)} \ell^\star(e)
\;=\;
(\kappa(P_{j-1})-1)-(\kappa(\widetilde P)-1)
\;=\;
d.
\]
Because $S\subseteq A_{\le j-1}\setminus E(G/\widetilde P)$ and every $e\in S$ has $\ell^\star(e)\ge L_{j-1}$,
\[
|S|\cdot L_{j-1}
\;\le\;
\sum_{e\in S}\ell^\star(e)
\;\le\;
d
\qquad\Longrightarrow\qquad
|S|
\;\le\;
\frac{d}{L_{j-1}}
\;=\;
d\,\lambda_{j-1}.
\]

\smallskip\noindent
\textbf{(2) Separating back to $P_{j-1}$ costs $|S|$ but yields $d$ extra components.}
Let $F^+ := F\cup S$ (i.e., additionally cut all edges of $S$). Since $S\subseteq A_{\le j-1}=E(G/P_{j-1})$,
after deleting $S$ there are \emph{no} remaining edges between distinct parts of $P_{j-1}$, hence
$\kappa(G-F^+) \ge \kappa(G-F)+d \ge k+d$.

\smallskip\noindent
\textbf{(3) Recovering cost by undoing $d$ internal splits.}
By~(2), we have $\kappa(G-F^+) \ge k+d$, so $G-F^+$ has $d$ \emph{surplus} components beyond $k$.
We will restore edges inside parts of $P_{j-1}$ to reduce the number of components by $d$ while
saving at least $\lambda_j$ edges per merge.

Fix any $C\in P_{j-1}$. Since $C$ is not split by the ideal-load recursion before level $j$,
we have $\Phi(G[C])\ge \lambda_j$ (by \cref{fact:ideal-loads-toolkit}(ii) and \cref{thm:psp-from-loads}).
Using $\Phi(G[C])\le \lambda(G[C])$ from \cref{fact:ideal-loads-toolkit}(i), it follows that
$\lambda(G[C])\ge \lambda_j$, and hence every non-trivial cut in $G[C]$ has size at least $\lambda_j$:
for all $\emptyset\subsetneq U\subsetneq C$,
\[
|\delta_{G[C]}(U)| \ge \lambda_j.
\]

Set $F^{(0)}:=F^+$. For $t=0,1,\dots,d-1$:
since $A_{\le j-1}\subseteq F^{(t)}$ (we only restore edges within a single $P_{j-1}$-part),
no edge remains between distinct parts of $P_{j-1}$ in $G-F^{(t)}$.
If every $C\in P_{j-1}$ induced a connected subgraph in $G[C]-F^{(t)}$, then
$\kappa(G-F^{(t)})\le \kappa(P_{j-1})<k$, contradicting $\kappa(G-F^{(t)})\ge k+(d-t)>k$.
Therefore, there exists some $C\in P_{j-1}$ such that $G[C]-F^{(t)}$ has at least two components.
Let $U\subsetneq C$ be the vertex set of any one component of $G[C]-F^{(t)}$ and define
\[
X_t := \delta_{G[C]}(U).
\]
Because $U$ is a connected component of $G[C]-F^{(t)}$, every edge of $X_t$ is cut by $F^{(t)}$, hence
$X_t\subseteq F^{(t)}$. Define
\[
F^{(t+1)} := F^{(t)} \setminus X_t.
\]
Restoring $X_t$ merges $U$ with $C\setminus U$, so
$\kappa(G-F^{(t+1)}) = \kappa(G-F^{(t)})-1$, and moreover
$|X_t|=|\delta_{G[C]}(U)|\ge \lambda_j$, hence
\[
|F^{(t+1)}| \le |F^{(t)}|-\lambda_j.
\]

After $d$ iterations let $\widehat F := F^{(d)}$. Then $\kappa(G-\widehat F)\ge k$ and
\[
|\widehat F|
\le
|F^+| - d\lambda_j
=
|F| + |S| - d\lambda_j
\le
|F| + d\lambda_{j-1} - d\lambda_j
<
|F|,
\]
using $|S|\le d\lambda_{j-1}$ from~(1) and $\lambda_{j-1}<\lambda_j$.
This contradicts the optimality of $F$. Hence $S=\emptyset$ and $A_{\le j-1}\subseteq F$.

Thus there exists an optimal $k$-cut that refines $P_{j-1}$, as claimed.
\end{proof}

%% file: 04_kt_kernels.tex
\section{Level-$\lambda_j$ KT Kernels}
\label{sec:kt}

\begin{definition}[Non-trivial cut]
A cut is \emph{non-trivial} if both sides contain at least two vertices.
\end{definition}

\begin{lemma}[KT kernel and static decomposition at threshold $\lambda_j$]\label{lem:kt}
For any part $C$ of $P_{j-1}$, there is a deterministic near-linear-time procedure that returns
\[
(K_C,\ \mathcal{D}_C,\ s_C)
\]
with the following properties:
\begin{enumerate}[label=(\roman*)]
\item $K_C$ is a capacitated kernel with $|V(K_C)|=\Otil(|C|/\lambda_j)$.
\item $\mathcal{D}_C$ is the \emph{KT decomposition forest} obtained by the cut-or-certificate recursion at threshold $\lambda_j$ (each internal edge corresponds to a non-trivial $\le\lambda_j$ cut in a sparse certificate, each leaf is a $\lambda_j$-core).
\item $s_C : V(K_C)\to\mathbb{Z}_{\ge1}$ is the \emph{core-size} map, where $s_C(x)$ equals the number of original vertices contracted into the supernode $x$.
\item (\emph{Preservation}) Every non-trivial cut in $G[C]$ of value $\le \lambda_j$ is a union of $\lambda_j$-cores and therefore appears in $K_C$ with the same value.
\end{enumerate}
The total time is $\Otil(|E(G[C])|+\lambda_j|C|)$ and the construction is deterministic.
\end{lemma}

\paragraph{Deterministic cut-or-certificate at threshold $\theta$.} When we invoke the cut-or-certificate algorithm, we actually pass the integer threshold $\lfloor\lambda_j\rfloor$. Since all cut values are integers, ``value $\leq \lambda_j$'' is equivalent to ``value $\leq \lfloor\lambda_j\rfloor$'', so the algorithm still finds and preserves exactly all non-trivial cuts of value at most $\lambda_j$.

\begin{proposition}[Cut-or-certificate, deterministic, \cite{KTACM}]\label{prop:cut-or-cert}
Given a simple graph $Q$ and $\theta$ a positive rational, there is a deterministic $\Otil(|E(Q)|)$ algorithm that either
\begin{enumerate}[label=(\alph*)]
\item outputs a non-trivial cut $(S,\overline S)$ with $|\delta_Q(S)|\le \theta$, or
\item certifies that every non-trivial cut in $Q$ has value $> \theta$,
\end{enumerate}
and, via repeated trimming of degree-$<\theta/2$ vertices, strengthens the certificate to ensure $\mindeg{Q}\ge \theta/2$ on the resulting piece.
\end{proposition}

\begin{remark}[Kernels preserve only non-trivial cuts]\label{rem:kernel-nontrivial}
Lemma~\ref{lem:kt} preserves \emph{non-trivial} internal cuts of value $\le \lambda_j$ in a part $C$. It does not preserve trivial (singleton) costs; hence any island routine must operate on $G[C]$, not on $K_C$.
\end{remark}

\begin{remark}[Parallel edges merged]
Contractions may create parallel edges; we merge them and carry capacities. All primitives operate correctly on the capacitated graph.
\end{remark}

\begin{definition}[Kernel views and restrictions]\label{def:view}
Let $C$ be an active part with kernel $(K_C,\mathcal{D}_C,s_C)$. For any descendant part $C'\subseteq C$ that is a union of $\lambda_j$-cores, the \emph{kernel view} of $C'$ is the induced subgraph $K_C[C']$ on the supernodes contained in $C'$; the \emph{restricted forest} is the induced subforest of $\mathcal{D}_C$ on those nodes, and the size map is $s_{C'}:=s_C|_{V(K_C[C'])}$.
\end{definition}

\begin{lemma}[Induced views under level-$\lambda_j$ borders: no KT recomputation]\label{lem:induced-view}
If a non-trivial border $B=(X,C\setminus X)$ of value $\le\lambda_j$ splits $C$ into $C_1=X$ and $C_2=C\setminus X$, then $C_1$ and $C_2$ are unions of $\lambda_j$-cores. Consequently,
\[
K_{C_1}=K_C[C_1],\qquad K_{C_2}=K_C[C_2],
\]
and the corresponding decomposition forests and size maps are obtained by pointer restriction from $(\mathcal{D}_C,s_C)$. No additional cut-or-certificate work is required for descendants of $C$.
\end{lemma}

\begin{proof}
By Lemma~\ref{lem:kt} every non-trivial $\le\lambda_j$ cut is a union of cores, so both $C_1$ and $C_2$ are unions of cores. The claims follow from Definition~\ref{def:view}.
\end{proof}

\begin{lemma}[Lift of kernel borders]\label{lem:lift}
Let $B$ be a border (non-trivial cut) chosen in a kernel view of $C$ with $\val(B)\le \lambda_j$. Then there is a canonical lift $B^{\uparrow}$ in $G[C]$ with $\val(B^{\uparrow})=\val(B)$ such that replacing $C$ by the two sides of $B^{\uparrow}$ refines the partition $\Pi$ consistently.
\end{lemma}

\begin{proof}
Let $\pi: V(G[C])\to V(K_C)$ send each vertex to its core’s supernode. If $B$ is $(S',\overline S')$ in the view, define $S:=\pi^{-1}(S')$ and set $B^\uparrow:=\delta_{G[C]}(S)$. Only intra-core edges were suppressed; crossing capacities are preserved, so $\val(B^\uparrow)=\val(B)$.
\end{proof}

\begin{lemma}[Global kernel size]\label{lem:global-kernel}
Let $\mathcal{K}:=\{K_C: C\in P_{j-1}\}$ be the level-$j$ kernels. Then
\[
\sum_{C\in P_{j-1}} |V(K_C)| \;=\; \Otil(n/\lambda_j),
\]
and this potential bound is preserved under any sequence of level-$\lambda_j$ internal splits by replacing a parent’s view with the two induced child views (Lemma~\ref{lem:induced-view}).
\end{lemma}

\begin{proof}
From Lemma~\ref{lem:kt}, $|V(K_C)|=O(|C|/\lambda_j)$ for each $C$, whence $\sum_C |V(K_C)|=\Otil(n/\lambda_j)$. Under a split $C\to C_1,C_2$, the child views are induced subgraphs on disjoint sets of supernodes from $K_C$, so $|V(K_{C_1})|+|V(K_{C_2})|\le |V(K_C)|$ up to $\poly(\log)$ factors.
\end{proof}

\subsection{Canonical Border Family at Level $\lambda_j$}
\label{subsec:border-family}

\begin{definition}[Border (internal refinement)]\label{def:border}
A border is a non-trivial cut of value $\le \lambda_j$ taken inside a current part (a descendant of $P_{j-1}$). Choosing a border increases the number of parts by one.
\end{definition}

\begin{remark}[Threshold equality at level $j$]
\label{rem:threshold_equality}
Inside parts of $P_{j-1}$, there are no non-trivial internal cuts of value $<\lambda_j$ by the definition of $\lambda_j$; hence every border used at level $j$ has value exactly $\lambda_j$.
\end{remark}

\paragraph{Frontier-based candidate enumeration (no GH trees).}
We implement \textsc{Build-Candidate-Borders} by scanning the \emph{frontier edges} of the static forests $\mathcal{D}_C$: these are recursion edges whose two child pieces both lie inside the current view of $C$. Each such edge corresponds to a certified non-trivial $\le\lambda_j$ cut between its two child pieces. We keep only those whose lift to $G[C]$ is non-trivial using the size map $s_C$.

\begin{algorithm}[H]
\caption{\textsc{Build-Candidate-Borders}$(\{(K_C,\mathcal{D}_C,s_C)\}, \lambda_j)$}
\label{alg:list-frontier}
\KwIn{For each active part $C$, its kernel view $K_C$, decomposition forest $\mathcal{D}_C$, and core-size map $s_C$}
\KwOut{A set $\mathcal{B}$ of lifted non-trivial borders of value $\le\lambda_j$}
$\mathcal{B}\gets\emptyset$\;
\ForEach{$C$ active}{
  \ForEach{frontier edge $e=(U,V)$ of $\mathcal{D}_C$ that is \emph{internal to} the current view of $C$}{
    \If{$\min\{\sum_{x\in U}s_C(x),\,\sum_{x\in V}s_C(x)\}\ge 2$}{
      Add the lifted cut $B^\uparrow(e)$ in $G[C]$ to $\mathcal{B}$\;
    }
  }
}
\Return $\mathcal{B}$\;
\end{algorithm}

\begin{lemma}[Coverage by a canonical border family]\label{lem:border-coverage}
There is a deterministic procedure \textsc{Build-Candidate-Borders} (Algorithm~\ref{alg:list-frontier}) that, given the current kernel views at threshold $\lambda_j$, returns a set $\mathcal{B}$ of candidate borders satisfying:
\begin{enumerate}[label=(\alph*)]
\item \textbf{Completeness.} If, along some optimal refinement path, a non-trivial internal cut $B^\star$ of value $\le \lambda_j$ is needed at this depth inside a current part $C$, then $\mathcal{B}$ contains a border (in the same part) whose application keeps the instance compatible with that optimal path.
\item \textbf{Size.} $|\mathcal{B}| \le \Otil\!\big(\sum_C |V(K_C)|\big) = \Otil(n/\lambda_j)$.
\item \textbf{Independence \& additivity.} Repeatedly peeling pendant pieces in $\mathcal{D}_C$ yields a subfamily $\mathcal{B}_{\mathrm{ind}}\subseteq \mathcal{B}$ whose lifted borders are edge-disjoint in their original parts. Consequently, for any subset $T \subseteq \mathcal{B}_{\mathrm{ind}}$,
$$\mathrm{val}\left(\bigcup_{B\in T}B\right) = \sum_{B\in T}\mathrm{val}\left(B\right) \leq |T| \cdot \lambda_j$$
\item \textbf{Non-triviality after lifting.} Every candidate lifts to a cut in $G$ whose two sides each contain at least two original vertices (by the size filter).
\end{enumerate}
\end{lemma}

%% file: 05_algorithm.tex
\section{Algorithmic Framework}
\label{sec:algo}

\paragraph{State.}
A state consists of: (i) a refinement $\Pi$ of $P_{j-1}$,
(ii) for every active part $C\in\Pi$ its \emph{level-$j$ kernel view} together with the static structures
$(K_C,\ \mathcal{D}_C,\ s_C)$ returned by Lemma~\ref{lem:kt}, and
(iii) the remaining split budget $R'$.
Kernel \emph{views} are induced subgraphs of an ancestor kernel; the forests $\mathcal{D}_C$ and size
maps $s_C$ are pointer restrictions (Definition~\ref{def:view}, Lemma~\ref{lem:induced-view}).

\subsection{Top-level driver}

\begin{algorithm}[H]
\caption{\textsc{PSP-KT-MinKCut}$(G, k)$}
\label{alg:driver}
\KwIn{Graph $G$, integer $k$}
\KwOut{A minimum $k$-cut}

Compute a PSP $(P_0 \succ \cdots \succ P_t)$ and strengths $(\lambda_1, \dots, \lambda_t)$\;
$j \gets \min\{\,i : \kappa(P_i) \ge k\,\}$\;
$R \gets k - \kappa(P_{j-1})$\;
$\Pi \gets P_{j-1}$\;

\ForEach{$C \in \Pi$}{
  $(K_C,\ \mathcal{D}_C,\ s_C) \gets \textsc{KT-Decompose}(C, \lambda_j)$\tcp*[r]{by Lemma~\ref{lem:kt}}
}

\Return $\textsc{DFS-Search}(\Pi, \{(K_C,\mathcal{D}_C,s_C)\}_{C\in\Pi}, R)$\;
\end{algorithm}

\subsection{DFS with canonical borders}
\begin{algorithm}[H]
\caption{\textsc{DFS-Search}$(\Pi, \{(K_C,\mathcal{D}_C,s_C)\}, R')$}
\label{alg:dfs}
\KwIn{Current partition $\Pi$, kernel views $\{(K_C,\mathcal{D}_C,s_C)\}$, and remaining budget $R'$}
\KwOut{Updated best cut configuration}

\If{$R' = 0$}{
  Assemble cut from $\Pi$; update best \\
  \Return
}

$\mathcal{B} \gets \textsc{Build-Candidate-Borders}(\{(K_C,\mathcal{D}_C,s_C)\}, \lambda_j)$\tcc*[r]{by Lemma~\ref{lem:border-coverage}; frontier edges of $\mathcal{D}_C$, filtered so each lift $B^\uparrow$ is non-trivial}

\ForEach{$B \in \mathcal{B}$}{
  Apply $B^\uparrow$ to its part $C$ (by Lemma~\ref{lem:lift}); obtain children $C_1, C_2$\;
  \tcp{No recomputation: induced views only (Lemma~\ref{lem:induced-view})}
  $K_{C_1} \gets K_C[C_1]$, $K_{C_2} \gets K_C[C_2]$\;
  restrict $\mathcal{D}_C$ to $\mathcal{D}_{C_1},\ \mathcal{D}_{C_2}$ and $s_C$ to $s_{C_1},\ s_{C_2}$;\label{line:viewupdate}
  \textsc{DFS-Search}$(\Pi', \{(K_{C'},\mathcal{D}_{C'},s_{C'})\}, R'-1)$\;
}

\tcp{Also consider stopping here and realizing the remaining $R'$ increases via islands, run on the current parts $\Pi$, not on kernels; see Remark~\ref{rem:kernel-nontrivial}.}
\textsc{Solve-Islands}$(\Pi, R')$\ \text{update best}\;
\end{algorithm}

\subsection{Islands (algebraic)}\label{subsec:islands}
\begin{proposition}[Exact island cost]\label{prop:island-cost-exact}
For any graph $H$ and $S\subseteq V(H)$, the number of edges that must be deleted
to isolate every vertex in $S$ equals $|\delta_H(S)|+|E_H(S)|$. Equivalently,
$\sum_{v\in S}\deg_H(v) - |E_H(S)|$.
\end{proposition}
\begin{proof}
Every edge with exactly one endpoint in $S$ must be removed, contributing $|\delta_H(S)|$.
Every edge with both endpoints in $S$ must also be removed, contributing $|E_H(S)|$.
Since $\sum_{v\in S}\deg_H(v) = |\delta_H(S)| + 2|E_H(S)|$, the two formulas coincide.
\end{proof}

\begin{lemma}[Island--border commutativity]\label{lem:island-border-commute}
Let $\Pi$ be any refinement of $P_{j-1}$. 
Fix a part $C\in\Pi$, a vertex $v\in C$, and a non-trivial internal border 
$B=(X,C\setminus X)$ in $G[C]$.
%with $\val(B)\le\lambda_j$.
Consider the two sequences of refinement operations inside $C$:
\begin{enumerate}[label=(\roman*)]
\item \textbf{Island-first:} isolate $v$ in $C$ (cost $\deg_{G[C]}(v)$) 
and then apply the restriction of $B$ to $C\setminus\{v\}$;
\item \textbf{Border-first:} apply $B$ in $C$ (cost $\val(B)$) 
and then isolate $v$ inside whichever side of $B$ contains $v$.
\end{enumerate}
Both sequences end in the same partition of $C$, namely 
$\{\{v\},\,X\setminus\{v\},\,C\setminus X\}$, and their total costs are equal.
Consequently, any schedule of singleton isolations and 
non-trivial borders of value $\le\lambda_j$ can be reordered so that
all islands are performed last without changing the final cost.
\end{lemma}

\begin{algorithm}[H]
\caption{\textsc{Solve-Islands}$(\Pi, r)$}
\label{alg:islands}
\KwIn{Current partition $\Pi$ (a refinement of $P_{j-1}$) and an integer $r$}
\KwOut{An $r$-island solution on $G$}
Let $M \leftarrow \sum_{C \in \Pi} |C|$ be the size of the current active parts (so $M \le n$).\\
Solve the exact $r$-islands subproblem on the disjoint union $\biguplus_{C\in\Pi} G[C]$:
   find $S\subseteq V$ with $|S|=r$ minimizing
   \[
      \mathrm{cut}(S) := |\delta_{\biguplus_{C\in\Pi} G[C]}(S)| \;+\; |E_{\biguplus_{C\in\Pi} G[C]}(S)|,
   \]
   equivalently $\mathrm{cut}(S)=\sum_{v\in S}\deg_{G[C(v)]}(v)\;-\;|E_{\biguplus_{C\in\Pi} G[C]}(S)|$.
   Use the algebraic routine (for example from \cite[Lemma 2.4]{HeLi22}\footnote{Note. Lemma 2.4 in \cite{HeLi22} gives a deterministic algorithm.}) in time $\Otil\!\left(M^{(\omega/3)\,r + O(1)}\right)$, where
   $M=\sum_{C\in\Pi} |C| \le n$.
\\
Return the corresponding cut in $G$.
\end{algorithm}
\noindent
\emph{Note.} By Remark~\ref{rem:kernel-nontrivial}, running on kernels would not preserve singleton costs. Operating on $\Pi$ ensures the value is exact and realizable in $G$.

\begin{remark}[Submodular form]
Let $f_C(S):=|\delta_{G[C]}(S)|$ denote the symmetric submodular cut function.
For $v\in X$,
\[
f_C(\{v\})+f_{C\setminus\{v\}}(X\setminus\{v\})
 = f_C(\{v\})+f_C(X)-|\delta_{G[C]}(v, C\setminus X)|
 = f_C(X)+|A|,
\]
which is identical to the border-first total above.
This view makes the commutativity a direct corollary of submodularity.
\end{remark}

\begin{corollary}[Deferred islands]\label{cor:stop-islands}
In the DFS of Algorithm~\ref{alg:dfs}, it is without loss of generality
to defer all singleton isolations to the leaf.
Along some optimal branch, every remaining increase in $\kappa$ 
can be realized as islands on the current parts~$\Pi$ 
with the same total value as any interleaved schedule.
This justifies the ``stop-and-island'' step in 
Line~9 of Algorithm~\ref{alg:dfs}
and the proof of Theorem~\ref{thm:correct}.
\end{corollary}

\subsection{Framework invariants}

\begin{lemma}[Kernel preservation]\label{lem:preserve}
Every non-trivial cut of value $\le \lambda_j$ inside any active part $C$ exists in its kernel \emph{view} $K_C$ with the same value.
\end{lemma}

\begin{proof}
Fix an active part $C$ and any non-trivial cut $S \subsetneq C$ with $|\delta_{G[C]}(S)| \le \lambda_j$.
By Lemma~\ref{lem:kt}, $C$ is decomposed into $\lambda_j$-cores with no non-trivial $\le\lambda_j$ cut; hence $S$ is a union of cores and is preserved under contraction.
Under any refinement by a non-trivial $\le\lambda_j$ border, children $C_1,C_2$ are unions of cores and inherit \emph{induced} views $K_{C_1}=K_C[C_1],K_{C_2}=K_C[C_2]$ (Lemma~\ref{lem:induced-view}; cf. Line~\ref{line:viewupdate}).
Thus the invariant holds inductively.
\end{proof}

\begin{lemma}[Global size invariant]\label{lem:size-inv}
At all recursion depths, $\sum_C |V(K_C)|=\Otil(n/\lambda_j)$.
\end{lemma}

\begin{proof}
Initially (parts of $P_{j-1}$), Lemma~\ref{lem:kt} gives $|V(K_C)|=\Otil(|C|/\lambda_j)$; summing yields $\Otil(n/\lambda_j)$.
Under a border, we replace one parent view by two \emph{induced} child views on a disjoint partition of the parent’s supernodes (Lemma~\ref{lem:induced-view}), so $|V(K_{C_1})|+|V(K_{C_2})|\le |V(K_C)|$.
%up to polylog factors.
Thus, the total $\sum_C|V(K_C)|$ never increases beyond its initial $\Otil(n/\lambda_j)$ bound.
\end{proof}

\begin{lemma}[Budget invariant]\label{lem:budget}
Each border branch reduces $R'$ by one; the island branch consumes the remaining $R'$ in one step.
\end{lemma}

\begin{proof}
A non-trivial border increases $\kappa(\Pi)$ by exactly one, so the recursive call uses $R'-1$.
When the island branch is taken, we realize $r=R'$ singletons at once and stop.
\end{proof}

\begin{remark}[Islands do not rely on kernels]
When the island branch is taken, \textsc{Solve-Islands} runs on the current parts $\Pi$.
Thus \cref{lem:preserve} (kernel preservation) is irrelevant to islands, and soundness follows
directly from operating on $G[C]$.
\end{remark}

%% file: 06_correctness.tex
\section{Correctness and Running Time}
\label{sec:crt}

\subsection{Correctness}
\begin{remark}[Enumerative completeness]\label{lem:complete}
By \Cref{lem:psp-normal-form}, there exists an optimal path that uses only level-$j$ borders and islands.
At each depth, \Cref{lem:border-coverage} ensures a correct border appears in $\mathcal{B}$; \Cref{lem:preserve}
guarantees it is present in the kernel \emph{views}. Hence some recursion branch realizes $\opt$.
\end{remark}

\begin{lemma}[Soundness of border application]\label{lem:sound}
Whenever $B\in\mathcal{B}$ is chosen, applying $B^\uparrow$ and then \emph{updating kernel views by induced restriction}
(Line~\ref{line:viewupdate}) preserves the invariants of \Cref{lem:preserve,lem:size-inv,lem:budget} and keeps the instance
compatible with an optimal continuation (by \Cref{lem:border-coverage}).
\end{lemma}

\begin{proof}
Fix any current state \sloppy$(\Pi,\{(K_C,\mathcal{D}_C,s_C)\},R')$ and any candidate border $B\in\mathcal{B}$ returned by
\textsc{Build-Candidate-Borders}. Let $C$ be the part of $\Pi$ in which $B$ lies. We check that applying $B^\uparrow$ and then
restricting to the two child \emph{views} at the same threshold $\lambda_j$ (Line~\ref{line:viewupdate}) preserves each framework
invariant, and that when $B$ is the witness promised by \Cref{lem:border-coverage}, the instance remains compatible with an
optimal continuation.

\smallskip
\emph{(i) Valid refinement \& value preservation.}
By \Cref{lem:lift}, lifting $B$ to $G[C]$ gives a cut $B^\uparrow$ with
$\val(B^\uparrow)=\val(B)\le \lambda_j$; replacing $C$ by the two sides of $B^\uparrow$ refines
$\Pi$ consistently.

\smallskip
\emph{(ii) Kernel preservation invariant (\Cref{lem:preserve}).}
After the split we do \emph{not} recompute KT; by \Cref{def:view,lem:induced-view}, the children $C_1,C_2$ inherit
\emph{induced} views $K_{C_1}=K_C[C_1]$, $K_{C_2}=K_C[C_2]$ and restricted forests/size maps. Every non-trivial $\le\lambda_j$ cut
inside $C_1$ or $C_2$ is present in the respective view with the same value (by \Cref{lem:kt}), so \Cref{lem:preserve} holds.

\smallskip
\emph{(iii) Global size invariant (\Cref{lem:size-inv}).}
By \Cref{lem:global-kernel} and \Cref{lem:induced-view}, replacing one parent view by two induced child views does not increase
$\Phi:=\sum_{C'} |V(K_{C'})|$ beyond $\Otil(n/\lambda_j)$.

\smallskip
\emph{(iv) Budget invariant (\Cref{lem:budget}).}
A border is non-trivial by definition, so it increases $\kappa(\Pi)$ by exactly $1$; the recursive
call passes $R'-1$ as required (Line~\ref{line:viewupdate}).

\smallskip
\emph{(v) Compatibility with an optimal continuation.}
If $B$ is the particular candidate guaranteed by \Cref{lem:border-coverage}(a), then $B^\uparrow$
is nested inside the next optimal internal refinement; applying it does not split any targeted
part in the fixed optimal $k$-partition and therefore keeps the instance compatible with that
optimal path.

\smallskip
Items (i)–(iv) hold for \emph{every} $B\in\mathcal{B}$; item (v) holds for the witness border
promised by \Cref{lem:border-coverage}(a).
\end{proof}

\begin{theorem}[Correctness]\label{thm:correct}
\textsc{PSP-KT-MinKCut} returns a minimum $k$-cut.
\end{theorem}

\subsection{Running time}

\begin{lemma}[Per-depth fan-out]\label{lem:fanout}
At any depth, $|\mathcal{B}| \le \Otil\!\big(\sum_C |V(K_C)|\big)=\Otil(n/\lambda_j)$ by \Cref{lem:border-coverage,lem:size-inv}.
Moreover, enumerating $\mathcal{B}$ from the forests' frontiers takes $\Otil(n/\lambda_j)$ pointer time.
\end{lemma}

\begin{proof}
Let $\Pi$ be the current refinement and $\{(K_C,\mathcal{D}_C,s_C)\}_{C\in\Pi}$ the active kernel \emph{views}.
By \Cref{lem:border-coverage}(b), each part $C$ contributes $O(|V(K_C)|)$ frontier candidates (after the non-triviality filter),
hence $|\mathcal B|\le \sum_{C\in\Pi} O(|V(K_C)|)=\Otil(n/\lambda_j)$ by \Cref{lem:size-inv}.
Since frontier edges and size checks are scanned via pointers in $\mathcal{D}_C$ and $s_C$, the enumeration work is $\Otil(n/\lambda_j)$.
\end{proof}

\begin{lemma}[Islands cost]\label{lem:islands-cost}
Realizing $r$ singletons on the current parts $\Pi$ costs $M^{(\omega/3)r+O(1)}$, where
$M := \sum_{C\in\Pi} |C| \le n$.
\end{lemma}

\begin{proof}
By Algorithm~\ref{alg:islands}, we solve the exact $r$-islands subproblem on $\biguplus_{C\in\Pi} G[C]$
in time $M^{(\omega/3)r+O(1)}$. Since $M \le n$, this is at most $n^{(\omega/3)r+O(1)}$.
\end{proof}

\begin{theorem}[Running time]\label{thm:runtime}
Let $R:=k-\kappa(P_{j-1})$. The algorithm runs in
\[
T(n,m,k)=\Otil\!\left(\underbrace{\poly(m)+\lambda_j n}_{\text{preprocessing}} \;+\; \left(\frac{n}{\lambda_j}+n^{\omega/3}\right)^{R}\right).
\]
\end{theorem}

%% file: 07_last_meta.tex
\section{Deterministic $n^{c k}$ via PSP$\times$KT and a Small–$\lambda$ Routine}
\label{sec:nck}

\begin{lemma}[Kernel size and preservation at an arbitrary threshold]\label{lem:any-theta-kernel}
Let $\Pi$ be any refinement of $P_{j-1}$ and let $\theta\in\mathbb{Z}_{>0}$. For every part $C\in\Pi$ there is a deterministic procedure that returns
\[
\big(K_C^{(\theta)},\ \mathcal{D}_C^{(\theta)},\ s_C^{(\theta)}\big)
\]
such that:
\begin{enumerate}[label=(\roman*)]
\item $|V(K_C^{(\theta)})|= \Otil(|C|/\theta)$;
\item (\emph{Preservation}) every \emph{non-trivial} cut of $G[C]$ of value $\le \theta$ is a union of $\theta$-cores and hence appears in $K_C^{(\theta)}$ with the same value;
\item $\mathcal{D}_C^{(\theta)}$ is the static KT decomposition forest at threshold $\theta$ and $s_C^{(\theta)}$ maps each core-supernode to its original size.
\end{enumerate}
Consequently, $\sum_{C\in\Pi} |V(K_C^{(\theta)})|=\Otil(n/\theta)$.
\end{lemma}
\begin{proof}
Identical to Lemma~\ref{lem:kt} with $\theta$ in place of $\lambda_j$: compute a $\theta$-sparse certificate; run deterministic cut-or-certificate at threshold $\theta$ with trimming; contract the resulting $\theta$-cores to form $K_C^{(\theta)}$ while storing the recursion forest $\mathcal{D}_C^{(\theta)}$ and the size map $s_C^{(\theta)}$. In simple graphs each core has $\Omega(\theta)$ vertices, giving the size bound; preservation follows as in Lemma~\ref{lem:kt}. \qedhere
\end{proof}

\begin{lemma}[Monotone coverage w.r.t.\ threshold]\label{lem:theta-covers-lambdaj}
Fix $j$ and $P_{j-1}$. For any $\theta\ge \lambda_j$, build $\big(K_C^{(\theta)},\mathcal{D}_C^{(\theta)},s_C^{(\theta)}\big)$ for all $C\in P_{j-1}$.
Let $\mathcal{B}_\theta$ be the family returned by \textsc{Build-Candidate-Borders} applied to these $\theta$-level structures: namely, the lifted borders obtained by scanning the \emph{frontier edges} of the forests $\mathcal{D}_C^{(\theta)}$ that are internal to the current view, filtered so that each lift to $G[C]$ is non-trivial (both sides $\ge 2$ original vertices).
Then along the optimal refinement guaranteed by \Cref{lem:psp-normal-form}, at every depth there exists a border in $\mathcal{B}_\theta$ compatible with the next optimal non-trivial step. Moreover, $|\mathcal{B}_\theta|=\Otil(n/\theta)$.
\end{lemma}

\begin{proof}
(\emph{Completeness}) By \Cref{lem:psp-normal-form}, the next optimal internal step inside some $C\in P_{j-1}$ has value $\le \lambda_j \le \theta$. In $K_C^{(\theta)}$, every $\le\theta$ cut is a union of $\theta$-cores (Lemma~\ref{lem:any-theta-kernel}). Consider the cut induced by the optimal step in the view; the frontier of $\mathcal{D}_C^{(\theta)}$ along that side contains a pendant child whose incident recursion edge defines a certified $\le\theta$ border nested in the target cut. The size filter via $s_C^{(\theta)}$ ensures its lift is non-trivial in $G[C]$.

(\emph{Size}) Each $\mathcal{D}_C^{(\theta)}$ has $\Otil(|V(K_C^{(\theta)})|)$ edges over a $\theta$-sparse certificate, and we retain only frontier edges internal to the current view and passing the non-triviality test. Summing over parts yields $|\mathcal{B}_\theta|=\Otil(n/\theta)$.

The proof mirrors \Cref{lem:border-coverage} at level $\theta$, observing that only $\theta\ge\lambda_j$ is needed for coverage. \qedhere
\end{proof}

\begin{theorem}[Deterministic meta–theorem]\label{thm:nck}
Suppose there exists a deterministic exact routine for minimum $k$-cut on simple graphs that runs in time $\lambda^{t k}\, n^{O(1)}$ for some constant $t\ge 1$. Then there is a deterministic algorithm that finds a minimum $k$-cut in time
\[
T(n,k)\;=\;O\!\big(n^{\,c k+O(1)}\big),\qquad
c\;=\;\max\!\left\{\frac{t}{t+1},\ \frac{\omega}{3}\right\}\;<\;1.
\]
\end{theorem}

\begin{proof}
Compute a constant-factor estimate $\tilde\lambda$ of $\lambda_k$ deterministically. Let $(P_0\succ \cdots \succ P_t)$ be the canonical PSP and $j:=\min\{i:\kappa(P_i)\ge k\}$, $R:=k-\kappa(P_{j-1})$.

\emph{Small $\lambda$.} If $\lambda_k\le n^{1/(t+1)}$, the assumed routine solves the instance in $n^{(t/(t+1))k+O(1)}$ time.

\emph{Large $\lambda$.} Otherwise set
\[
\theta\;:=\;\max\{\lambda_j,\ \tilde\lambda\}.
\]
Build the $\theta$-level kernels and forests $\big(K_C^{(\theta)},\mathcal{D}_C^{(\theta)},s_C^{(\theta)}\big)$ for $C\in P_{j-1}$ once, and at each depth enumerate the candidate family $\mathcal{B}_\theta$ via \textsc{Build-Candidate-Borders} (frontier scan with non-triviality filter). By Lemma~\ref{lem:any-theta-kernel},
\[
\sum_{C\in P_{j-1}} |V(K_C^{(\theta)})|=\Otil(n/\theta),
\]
so per depth $A:=|\mathcal{B}_\theta|=\Otil(n/\theta)$. By Lemma~\ref{lem:theta-covers-lambdaj}, along the optimal path of \Cref{lem:psp-normal-form} there is always a compatible border in $\mathcal{B}_\theta$; by \Cref{cor:stop-islands} we may defer all singletons to the end. With $B:=n^{\omega/3}$ for the (exact) island call at the stop node,
\[
\sum_{\ell=0}^{R} A^{\ell} B^{\,R-\ell}\ \le\ (R{+}1)\cdot \max\{A,B\}^{\,R}.
\]
Since $R\le k$ and $\theta\ge \tilde\lambda\in\Omega\!\big(n^{1/(t+1)}\big)$ in this branch,
\[
A=\Otil(n/\theta)\ \le\ n^{t/(t+1)},\qquad
\max\{A,B\}\ \le\ n^{\max\{t/(t+1),\,\omega/3\}},
\]
so the large‑$\lambda$ branch runs in $\Otil\!\big(n^{\,\max\{t/(t+1),\,\omega/3\}\cdot k}\big)$. Combining the two branches proves the theorem.
\end{proof}

\begin{remark}[Instantiation with \cite{LSS20}]
Lokshtanov, Saurabh, and Surianarayanan~\cite{LSS20} give a parameterized approximation scheme for \emph{min $k$-cut}. Plugging such a routine into \Cref{thm:nck} yields a deterministic  $O(n^{c k+O(1)})$ algorithm for some constant $c<1$.
\end{remark}

\medskip
\noindent\emph{Note.} Along any root-to-leaf path we invoke \textsc{Solve-Islands} at most once (upon taking the stop branch), so the number of algebraic calls equals the number of leaves; the binomial upper bound $\big(\tfrac{n}{\lambda_j}+n^{\omega/3}\big)^R$ from Section~\ref{sec:crt} still applies. The enumeration and view updates at each node are pointer operations over the static $\theta$-forests; there is no per-node GH or KT recomputation.

%% file: omitted_proofs.tex
\subsection{Proofs from \Cref{sec:kt}}

\begin{proof}[Proof of \cref{lem:kt}]
\textbf{Step 1 (sparse certificate up to $\lambda_j$).}
Compute a $\lambda_j$-sparse certificate $H$ of $G[C]$ in time $\Otil(|E(G[C])|)$:
$H$ has at most $\lambda_j(|C|-1)$ edges and preserves all cuts of size $\le \lambda_j$.

\textbf{Step 2 (cut-or-certificate recursion).}
Recursively process induced pieces $Q\subseteq H$. For each piece, run the deterministic cut-or-certificate at threshold $\lambda_j$ (Proposition~\ref{prop:cut-or-cert}). If a non-trivial cut of value $\le\lambda_j$ is found, split $Q$ into its two sides; otherwise record $Q$ as a $\lambda_j$-core. Interleave the standard degree-$<\lambda_j/2$ trimming so that each certified core $X$ satisfies $\mindeg{X}\ge \lambda_j/2$.

\textbf{Step 3 (contraction and bookkeeping).}
Contract each core $X$ to a supernode and merge parallel edges (carrying capacities) to form $K_C$; define $\mathcal{D}_C$ to be the recursion forest from Step~2 and $s_C(x):=|X|$ for the supernode $x$ created from $X$.

\textbf{Correctness \& bounds.}
Any non-trivial $\le\lambda_j$ cut cannot split a certified core, so it is a union of cores and is preserved by contraction with the same value. Since the graph is simple and $\mindeg{X}\ge\lambda_j/2$, each core has size $|X|\ge \lambda_j/2+1$, hence the number of cores and thus $|V(K_C)|$ is $O(|C|/\lambda_j)$. Each edge of $H$ participates in $O(\log n)$ recursion steps, giving total $\Otil(|E(H)|)=\Otil(\lambda_j|C|)$ time.
\end{proof}

\begin{proof}[Proof of Lemma~\ref{lem:border-coverage}]
Fix $j$ and the current refinement $\Pi$ of $P_{j-1}$.  
For each active part $C\in\Pi$, let $(K_C,D_C,s_C)$ denote the level-$\lambda_j$ kernel, static decomposition forest, and core-size map from Lemma~\ref{lem:kt}.  
Write $\theta := \lambda_j$.  
The procedure \textsc{Build-Candidate-Borders} (Algorithm~\ref{alg:list-frontier}) scans the frontier edges $e=(U,V)$ of $D_C$ whose endpoints remain inside the current view of $C$, and keeps $e$ iff
\[
\min\Big\{\sum_{x\in U}s_C(x),\ \sum_{x\in V}s_C(x)\Big\}\ \ge\ 2,
\]
adding the lifted cut $B^{\uparrow}(e)$ in $G[C]$ as defined in Lemma~\ref{lem:lift}.  
We verify items (a)–(d).

\paragraph{(a) Completeness.}
Let $B^\star=(X,C\setminus X)$ be the next non-trivial internal cut of value $\le\theta$ on a fixed optimal refinement path, taken inside some active part $C$.  
By Lemma~\ref{lem:kt}, $B^\star$ is preserved in the kernel $K_C$ with the same value, and $X$ is a union of $\theta$-cores.  
Let $\mathcal{E}_X$ denote the set of edges of $D_C$ crossing between $X$ and $C\setminus X$ in the forest.  
Deleting $\mathcal{E}_X$ partitions $D_C$ into components corresponding to these two sides.  
Choose the edge $e^\circ=(U^\circ,V^\circ)\in\mathcal{E}_X$ that was created \emph{earliest} in the recursion (i.e., the highest-level certified split nested in $B^\star$).  
By Proposition~\ref{prop:cut-or-cert}, that split was certified as a non-trivial $\le\theta$ cut in the certificate graph, so both sides contained at least two original vertices at the time of creation; hence
\[
\min\Big\{\sum_{x\in U^\circ}s_C(x),\ \sum_{x\in V^\circ}s_C(x)\Big\}\ \ge\ 2.
\]
Thus $e^\circ$ passes the size filter, and its lifted border $B^{\uparrow}(e^\circ)$ belongs to~$\mathcal{B}$.  
Since $U^\circ\subseteq X$ and $V^\circ\subseteq C\setminus X$, the lifted cut is nested inside $B^\star$ and hence compatible with the optimal continuation. 

\smallskip
\paragraph{(b) Size.}
Each forest $D_C$ has $O(|V(K_C)|)$ edges, since each split in the deterministic cut-or-certificate recursion of Lemma~\ref{lem:kt} contributes one edge.  
Scanning the frontier and applying the filter yields at most $O(|V(K_C)|)$ candidates per $C$.  
Summing over parts and invoking Lemma~\ref{lem:global-kernel} gives
\[
|\mathcal{B}| \;\le\; \sum_{C\in\Pi}O(|V(K_C)|)
\;=\; \Otil(n/\lambda_j).
\]

\smallskip
\paragraph{(c) Independence and additivity (pendant peeling).}
For each $C$, associate to every recursion edge $e=(U,V)$ the edge set
\[
\Delta_H(e):= \bigl\{\,\text{certificate edges of the $\theta$-sparse $H\subseteq G[C]$ whose endpoints are first separated by $e$}\,\bigr\}.
\]
Each edge of $H$ is charged to the earliest split that separates its endpoints, so the sets $\{\Delta_H(e)\}$ are disjoint.  
If $e$ passes the filter, the lifted border $B^{\uparrow}(e)$ satisfies
\[
\val_{G[C]}(B^{\uparrow}(e))\ =\ |\Delta_H(e)|\ \le\ \theta,
\]
since $H$ preserves all $\le\theta$ cuts.  
Peeling pendant frontier edges one by one (always taking a leaf in $D_C$) yields a subfamily $\mathcal{B}_{\mathrm{ind}}\subseteq\mathcal{B}$ whose supports in $G[C]$ are edge-disjoint; hence their values add: any $r$ such borders increase $\kappa$ by $r$ (one per non-trivial cut), and since their support edge sets $\Delta_H(e)$ are disjoint, the union value equals the sum of their individual values, which is at most $r\theta$.  
Because $\le\theta$ cuts never split cores (Lemma~\ref{lem:kt}), this process is well defined.

\smallskip
\paragraph{(d) Non-triviality after lifting.}
By construction, only edges $e=(U,V)$ passing the side-size test are kept.  
Let $\pi:V(G[C])\to V(K_C)$ be the contraction map, and $X_U=\pi^{-1}(U)$, $X_V=\pi^{-1}(V)$.  
Then $\min\{|X_U|,|X_V|\}\ge2$, so $B^{\uparrow}(e)=\delta_{G[C]}(X_U)$ is a non-trivial cut in $G[C]$.  
This exactly enforces the size filter of Algorithm~\ref{alg:list-frontier}.  
\hfill{\textit{(cf.\ Lemma~\ref{lem:lift})}}

\smallskip

All properties (a)–(d) follow.  
Therefore, \textsc{Build-Candidate-Borders} deterministically returns a canonical border family of size $\Otil(n/\lambda_j)$ that covers every optimal refinement choice at this level.
\end{proof}

\subsection{Proofs from \Cref{sec:algo}}
\begin{proof}[Proof of \cref{lem:island-border-commute}]
If $B$ lies in a different part $C'\neq C$, the operations are edge-disjoint and trivially commute.
Assume $B$ lies in $C$ and $v\in X$. 
Partition the edges incident to $v$ inside $C$ into
\[
A := \{(v,u): u\in X\setminus\{v\}\}, 
\qquad
B_v := \{(v,u): u\in C\setminus X\}.
\]
Then $\deg_{G[C]}(v)=|A|+|B_v|$ and 
$\val(B)=|\delta_{G[C]}(X)|$, which already counts all edges in $B_v$.

\smallskip
\noindent\emph{Island-first.}
Isolating $v$ removes $A\cup B_v$.  
The subsequent border on $C\setminus\{v\}$ costs
$|\delta_{G[C]}(X)|-|B_v|$.
Hence the total cost is
\[
\deg_{G[C]}(v)+(|\delta_{G[C]}(X)|-|B_v|)=|\delta_{G[C]}(X)|+|A|.
\]

\smallskip
\noindent\emph{Border-first.}
Applying $B$ costs $|\delta_{G[C]}(X)|$, and isolating $v$ afterwards 
removes the remaining edges $A$.
Thus the total cost is again $|\delta_{G[C]}(X)|+|A|$.

\smallskip
Since both totals coincide and the final partition is identical, 
the two operations commute exactly.
By repeatedly swapping adjacent (island, border) pairs, 
one can reorder any sequence so that all islands are applied last.
\end{proof}

\subsection{Proofs from \Cref{sec:crt}}
\begin{proof}[Proof of Theorem~\ref{thm:correct} (Correctness)]
By \Cref{lem:complete}, there exists an optimal refinement path that uses only level-$j$ borders and islands.
At each depth, $\mathcal{B}$ contains a border whose application is compatible with that path (\Cref{lem:border-coverage}(a)).
When we apply that border and update the two child \emph{views}, \Cref{lem:sound} ensures that (i) all invariants remain true
(so the state is well-formed for further recursion), and (ii) the instance remains compatible with the same optimal continuation.

Proceeding inductively, either: (a) we reach $R'=0$ and assembling the cut from $\Pi$ yields a feasible $k$-cut; or (b) when the
“all remaining splits are singletons” condition holds, \textsc{Solve-Islands}$(\Pi, R')$ runs on the current parts and returns the exact
minimum cost of isolating $R'$ vertices in $G$. Along the optimal branch, border values are preserved by lifting (\Cref{lem:lift}) and
all subsequent $\le\lambda_j$ internal choices remain available in the inherited views (\Cref{lem:kt,lem:induced-view}). Since the DFS
explores all candidates in $\mathcal{B}$ and retains the minimum among leaves, the returned solution equals $\opt$.
\end{proof}

\begin{proof}[Proof of \cref{thm:runtime}]
\emph{Preprocessing.}
Computing the PSP is $\poly(m)$ (which we do not optimize here). For each $C\in P_{j-1}$, running \textsc{KT-Decompose}$(C,\lambda_j)$ (sparse certificate
$\to$ cut-or-certificate recursion $\to$ contraction), returns $(K_C,\mathcal{D}_C,s_C)$ in total time $\Otil(\lambda_j n)$ and
size $\sum_C |V(K_C)|=\Otil(n/\lambda_j)$ (\Cref{lem:kt,lem:global-kernel}).

\emph{DFS recurrence.}
Write $A:=\Otil(n/\lambda_j)$ and $B:=n^{\omega/3}$. At any non-leaf node with budget $R'>0$, by
\Cref{lem:fanout} we (i) enumerate at most $A$ candidates in $\Otil(A)$ pointer time and (ii) recurse on each child with budget $R'-1$,
after updating views by induced restriction (Line~\ref{line:viewupdate}), which is $O(1)$ pointer work per child
(\Cref{lem:induced-view}). If instead the “all remaining splits are singletons” condition holds, we stop and call \textsc{Solve-Islands} once;
by \Cref{lem:islands-cost}, trying all $r\in\{0,\ldots,R'\}$ costs at most
\[
\sum_{r=0}^{R'} B^{\,r+O(1)}\;=\;(1+B)^{R'}\cdot\poly(n)\;\le\;(A+B)^{R'}\cdot\poly(n),
\]
since $A\ge 1$ and polylog factors are absorbed by $\Otil(\cdot)$. Thus the worst-case recurrence is
\[
T(R')\ \le\ \Otil\!\big(A + A\cdot T(R'\!-\!1)\big)\quad\text{or}\quad
T(R')\ \le\ \Otil\!\big((A{+}B)^{R'}\big),
\]
and in all cases $T(R')\le \Otil\!\big((A{+}B)^{R'}\big)$ by induction with base $T(0)=\Otil(1)$.

\emph{Putting it together.}
With $R'=R$ at the root and adding the $\Otil(m+\lambda_j n)$ preprocessing,
\[
T(n,m,k)\ \le\ \Otil\!\Big(\poly(m)+\lambda_j n\;+\;(A+B)^{R}\Big)
\ =\ \Otil\!\Big(\poly(m)+\lambda_j n\;+\;\big(\tfrac{n}{\lambda_j}+n^{\omega/3}\big)^{R}\Big).
\]
\end{proof}